# Deep Learning-driven Community Resilience Rating based on Intertwined Socio-Technical Systems Features

Kai Yin[1], Bo Li[1*], Ali Mostafavi[1]

[1] Zachry Department of Civil and Environmental Engineering, Texas A&M University, College Station, United States.

[*] Corresponding author: Bo Li, libo@tamu.edu.

These authors contributed equally: Kai Yin, Bo Li.

**Abstract**

Community resilience is a complex, multi-dimensional phenomenon that emerges from nonlinear interactions among diverse features within socio-technical systems. However, most existing assessment methods rely on index-based approaches that inadequately capture the heterogeneity of these systems and their dynamic interdependencies in shaping key resilience components, namely robustness, redundancy, and resourcefulness. To address this gap, this study introduces *Resili-Net*, an integrated three-module deep learning framework for rating community resilience. The model incorporates twelve measurable features across infrastructure, facility, and social systems, using publicly available data from multiple metropolitan statistical areas in the United States. *Resili-Ne*t classifies spatial areas into five distinct resilience levels, and its interpretability allows for identifying the key factors influencing resilience across communities. Scenario analyses simulate how changes in socio-technical conditions affect resilience levels, and a combined resilience–risk analysis reveals communities facing both high hazard risk and low resilience. These findings highlight priority areas for targeted interventions. By modeling resilience as an emergent property of interacting community features, Resili-Net offers a scalable, data-driven approach for resilience assessment that advances current practices through the use of machine learning and heterogeneous urban data.

## Introduction

Natural hazards pose a significant threat to cities and communities worldwide. As population growth and urbanization continue, an increasing number of people and assets will be exposed to these hazards [1]. To cope with such hazards, assessing the resilience of cities and communities becomes essential for developing effective strategies. Resilience refers to the capacity of a system to prepare for, adapt to, and recover from disruptions events [2].

Decades of research have assessed urban resilience against natural hazards from a variety of disciplinary perspectives. Many studies focus on individual urban systems, such as transportation networks [3-6], power grids [7-9], communication infrastructure and social systems [10, 11] to evaluate how the system withstands and recovers from disruption. While these system-specific analyses offer valuable insights, they often fall short in capturing the broader, interconnected nature of urban resilience.

Recognizing that cities function as complex systems, studies have moved beyond isolated system assessment to incorporate this complexity into resilience assessment. Framework such as PEOPLES [12] has been developed to integrate social, environmental, infrastructural, and institutional dimensions to evaluate community resilience. Similarly, Datola [13] identifies five key dimension including economy, society, environment, nature, and governance and nine capacities essential for assessing urban resilience in a holistic way. Helmrich, Kuhn [14] evaluate urban infrastructure systems from a social-ecological-technological perspective, offering a convergence-based approach to uncover interdependencies, feedback loops, and disruption pathways that influence urban resilience. Together, these efforts underscore the growing recognition that resilience must be approached through a complex, multidimensional lens.

However, operationalizing this complexity in a spatially explicit, scalable, and context-sensitive manner remains an open challenge. A common method for assessing community resilience involves selecting a set of indicators, assigning weights to each, and aggregating them into a composite index or synthetic score. For example, studies by Moghadas, Asadzadeh [15] assessed urban flood resilience in Tehran using a composite index built from six dimensions, with indicator weights assigned through the Analytic Hierarchy Process (AHP) and aggregated using TOPSIS to rank districts. Similarly, Anelli, Tajani [16] proposed a synthetic natural risk index for Rome, combining 23 indicators across hazard, exposure, and vulnerability dimensions using AHP-based weighting. While these methods provide structured assessments, they typically assume that indicators function independently and rely heavily on expert-based weighting, limiting their ability to reflect the interconnected nature of real-world urban systems. Some studies, such as those by Parizi, Taleai [17] and Parizi, Taleai [18], have taken a step further by applying methods like Interpretive Structural Modeling (ISM) and DEMATEL to explore relationships and structures among indicators. Although these techniques offer insights into indicator interdependencies, the relationships they identify are static and derived from expert input, providing only a partial representation of the complex, nonlinear interactions that characterize socio-technical systems. These limitations highlight the need for more integrated, systems-based approaches that go beyond indicator aggregation to capture the emergent, adaptive, and interdependent nature of community resilience.

Recent advances in deep learning have demonstrated strong potential for modeling complex, high-dimensional, and nonlinear relationships in urban systems. Emerging studies have begun to leverage deep learning in the context of urban hazard and resilience. For example, Yin and Mostafavi [19] proposed *FloodRisk-Net*, an unsupervised graph deep learning model for emergent flood risk profiling. Siam et al. [20] employed a hybrid deep neural network and fuzzy AHP model for national-scale flood risk mapping. These efforts show how deep learning can be used to capture hidden patterns in urban risk contexts. Additionally, unsupervised deep learning has been effectively used to identify latent risk patterns and system vulnerabilities in other domains such as healthcare [21], transportation [6], and infrastructure diagnostics [22], offering transferable methodologies for urban resilience analysis.

Building upon these identified gaps and the emerging potential of deep learning to model complexity in urban systems, this study presents *Resili-Net*, a deep learning-based framework for rating community resilience. Specifically, we introduce an integrated three-layer unsupervised deep learning framework designed to characterize the resilience profiles of urban communities by leveraging urban big data and machine intelligence. We apply the *Resili-Net* model to four of the most populous metropolitan statistical areas (MSAs) in the United States: Los Angeles, Chicago, Dallas, and Houston, which are selected due to their large populations and the heightened socioeconomic impacts associated with hazards in densely populated regions. The model operates at the spatial grid cell level, assigning resilience scores based on each cell's underlying features. By mapping these resilience ratings alongside flood risk data, the framework enables identification of resilience–risk profiles within each city, highlighting high-risk, low-resilience zones that require prioritized intervention. While demonstrated in large MSAs, the *Resili-Net* framework has the potential to generalize and adapt to cities of varying population sizes and geographic contexts.

This study advances the field of community resilience assessment through the development and application of a novel deep learning-based framework, *Resili-Net*. The key contributions of our work are threefold. First, this study introduces an integrated three-module unsupervised deep learning architecture that automatically learns latent representations of socio-technical system features. By capturing hidden structures and complex, non-linear interdependencies among socio-technical system features, *Resili-Net* overcomes key limitations of traditional indicator-based models, which often rely on expert-defined weights and assume linear, additive relationships. The model clusters spatial units into distinct resilience groups based on their emergent characteristics, enabling a more nuanced, data-driven understanding of urban resilience that is both scalable and adaptable to diverse urban contexts. Second, this study enhances the interpretability of the deep learning framework by incorporating a feature importance analysis using a random forest classifier. This step reveals the relative contribution of each resilience feature to the classification of spatial units, translating the model's abstract representations into actionable insights. By doing so, it provides a transparent basis for stakeholders and policymakers to understand which factors most influence resilience outcomes, enabling more targeted, data-informed decision-making. Third, the implementation of the proposed framework across multiple metropolitans demonstrates the practical utility of *Resili-Net*, examining how varying urban development scenarios influence resilience distributions. By integrating high-resolution flood risk data, the framework enables the identification of compound vulnerable zones which are simultaneously highly exposed to hazards and structurally low in resilience. This integrated resilience–risk profiling supports spatial prioritization for targeted interventions. Our approach uniquely combines unsupervised deep learning with interpretable modeling and scenario analysis, offering a more robust and scalable tool for cross-sectoral planning and strategic resilience enhancement.

## Results

### Three-layer unsupervised deep learning model for community resilience rating

We constructed an integrated three-module deep learning model (*Resili-Net*) for community resilience rating (Fig.1) based on 12 features related to the three components of resilience: robustness, redundancy, and resourcefulness using publicly available urban datasets. Table 1 in the Methods section summarizes the features used in the model. The description and computation of features, as well as the rationale for their selection, are explained in the Methods section.

As presented in Fig. 1, in the first module, node representation learning, the complex and nonlinear interactions of resilience-related features (***RF***, Table 2) among different urban sub-systems are encoded into latent representations. In the second module, grid cells with similar resilience levels are clustered into the same group by implementing deep embedded clustering on the learned node representation. In the third module, feature importance is determined by utilizing clustering results as labels for classification algorithms and then interpreting cluster assignments modeled by the classifier. The community resilience level of each cluster is calculated by taking feature importance as weights.

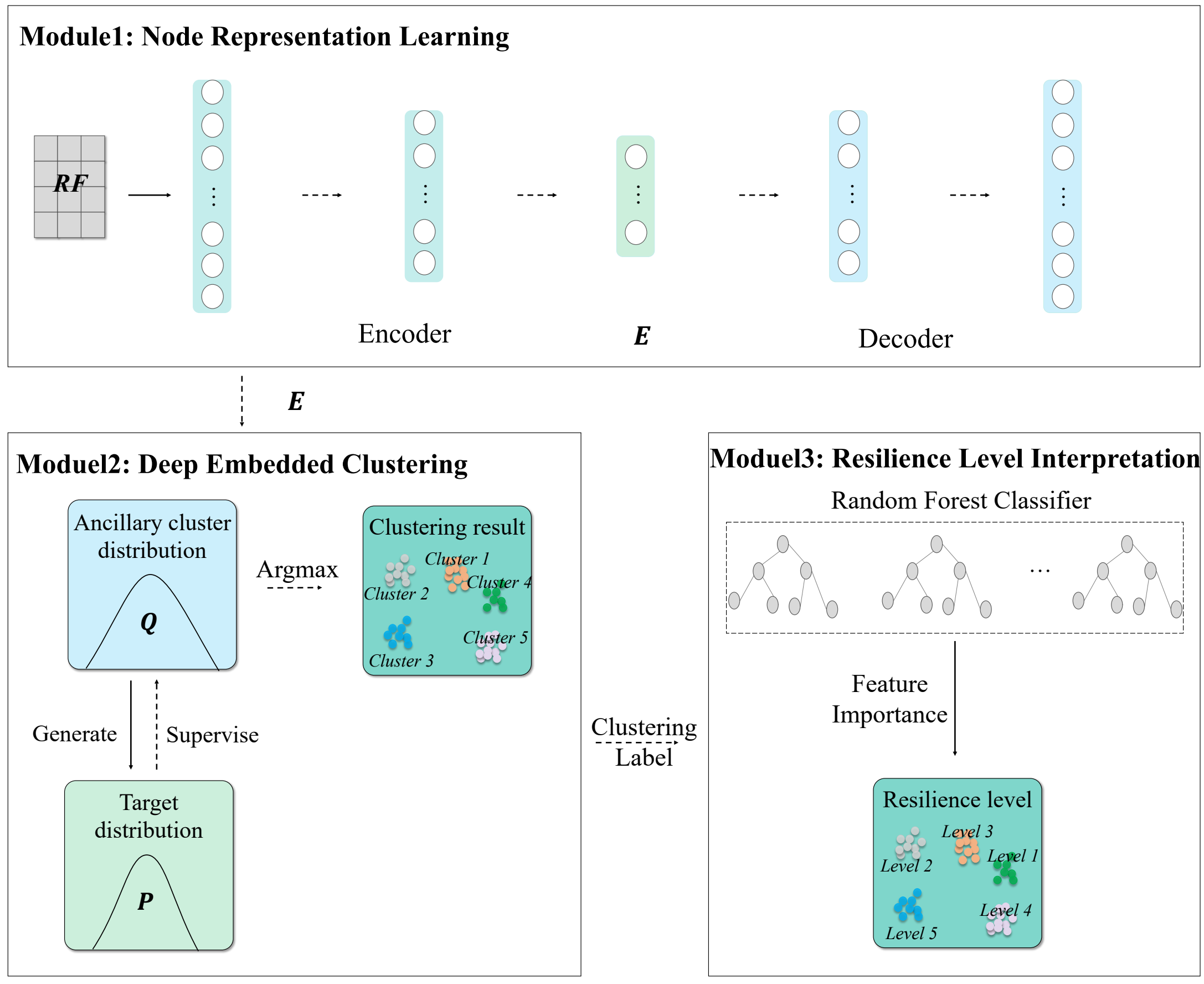

**Fig. 1 | The overall framework of the proposed three-module community resilience rating model.** Each MSA is divided into grid cells of equal size. The input of the model is resilience-related features matrix ($RF$) which includes three resilience characteristics of robustness, redundancy, and resourcefulness in urban sub-systems of facilities, infrastructures, and societies (Table 2; See the Methods for details about the construction of $RF$). The model outputs the resilience level (level 5 is the highest resilience, while level 1 is the lowest) of each grid cell. $E$ is the encoded representation of $RF$.

### Revealing five resilience levels within each MSA based on interactions among community resilience-related features

*Resili-Net* yields five distinct, city-specific community resilience levels, with level 5 being the highest resilience, while level 1 being the lowest resilience level (Fig. 2). These outputs unveil the spatial profile of community resilience across different areas of cities in an interpretable way. Each resilience level has distinct feature values related to the extent of robustness, redundancy, and resourcefulness of the urban sub-systems (Fig. 3). For example, the highest resilience level areas in Greater Houston are found to have the lowest building age, cell tower age, and poverty rate as well as the largest available number of healthcare centers within 30 minutes driving distance, the largest number of connected roads to the community, the largest available number of cell towers, the largest road length, the fastest internet speed, and the highest education level. All these features indicate that these communities have the greatest levels of robustness, redundancy, and resourcefulness, giving them the highest capacity to prepare for, absorb, recover from, and adapt to disturbances caused by hazards. The spatial distribution maps of the individual features for the four MSAs are provided in the Supplementary Figure 1 to 4.

To explore broader spatial patterns, we overlaid each MSA's core city boundary with its resilience mapping. Result show that higher-resilience communities tend to cluster in or near core cities across all four MSAs (Fig.2). These communities are characterized by the presence of newer and more robust infrastructure and facilities access, as well as higher educational levels (Fig.3). We also found that community resilience levels are spatially concentrated instead of randomly distributed and dispersed. Global Moran's I index is calculated by taking a contiguous network (grid cell sharing a boundary or a vertex is defined as a neighbor) as spatial-weight matrix. The significantly positive Moran's I (Table 1) indicates that grid cells with similar resilience are spatially clustered. Such clustering may arise because adjacent communities share infrastructure networks, socio-demographic characteristics, and exposure to common hazards; as a result, improvements made in one cell could propagate benefits (or conversely burdens) to nearby cells. It may also reflect past planning choices that have unevenly allocated resilience-enhancing resources. Determining which mechanism dominates will require targeted causal analysis beyond the scope of this study. As a comparison, we compiled resilience maps generated by a linear aggregation of the same features and simply classified into quintiles for all four MSAs (Supplementary Figure 5). We find that our deep-learning approach preserves sharper gradients and reveals spatial heterogeneities (such as central vs. peri-urban contrasts) that the quintile method smooths over, indicating its superior capacity to capture non-linear indicator interactions.

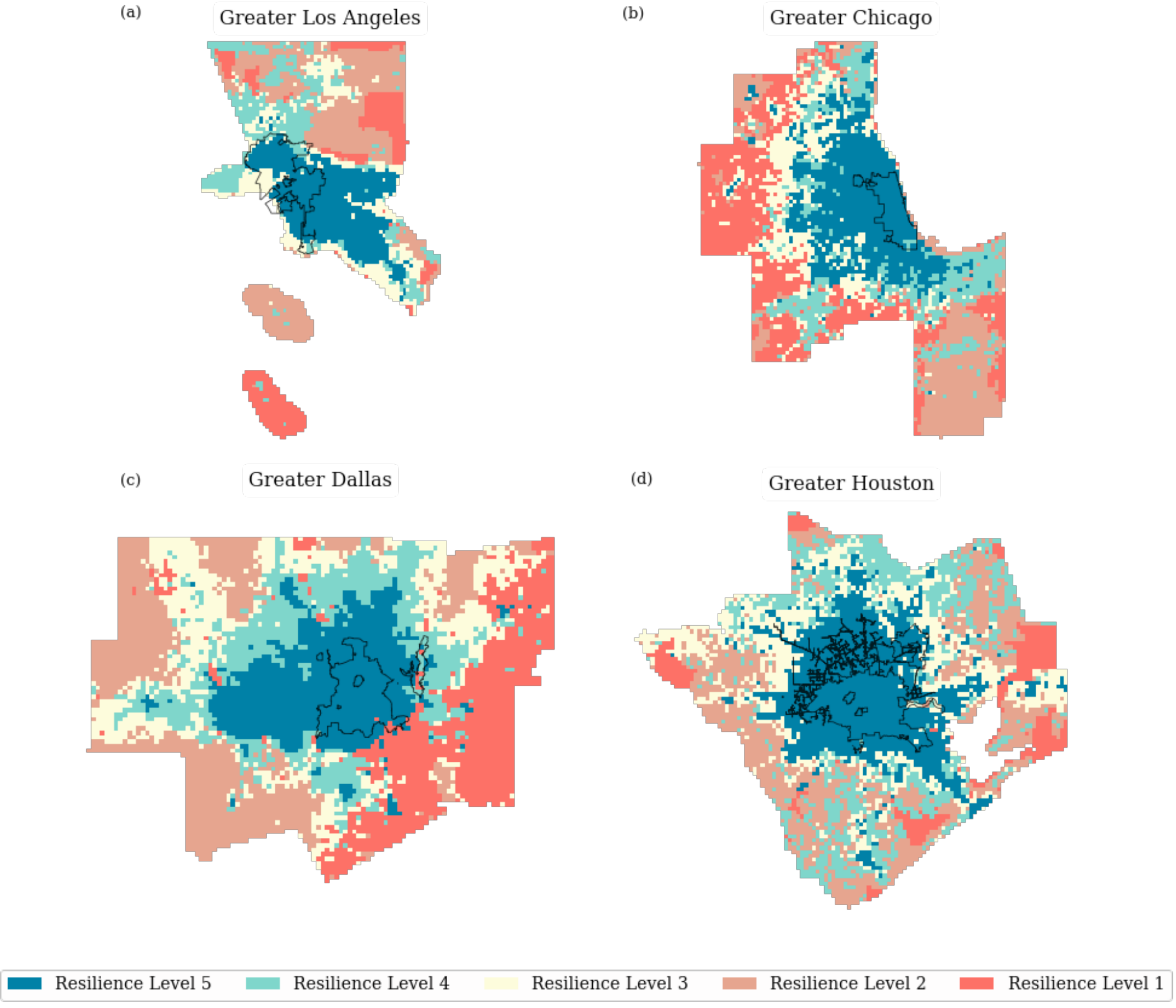

**Fig.2 | Geographical distribution of community resilience level in four MSAs.** Geographical distribution of community resilience levels in (a) Greater Los Angeles, (b) Greater Chicago, (c) Greater Dallas, (d) Greater Houston. Level 5 represents the highest resilience, while level 1 represents the lowest resilience level. The black lines in the maps represent the core cities within each MSA.

**Table 1 Results of Global Moran's I index in four MSAs**

| MSAs | Greater Los Angeles | Greater Chicago | Greater Dallas | Greater Houston |
|---|---|---|---|---|
| Moran's I value | 0.851*** | 0.768*** | 0.821*** | 0.698*** |

*** significant at 0.01 level

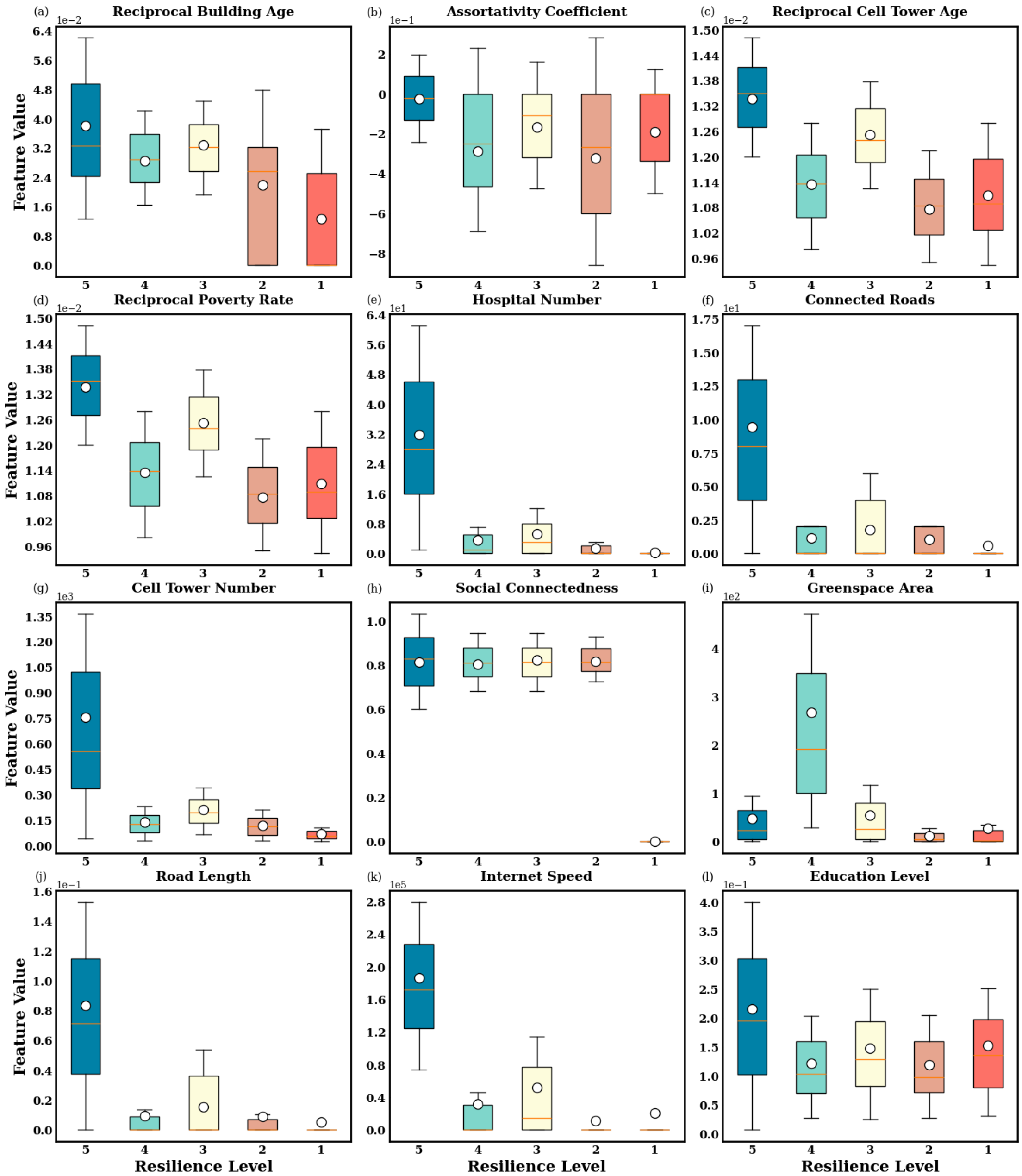

**Fig.3 | Characteristics of clusters in Greater Houston**. Community resilience comprises the features of (a) reciprocal building age, (b) assortativity coefficient, (c) reciprocal cell tower age, (d) the reciprocal rate of poverty, (e) the number of healthcare centers within 30 minutes driving distance, (f) the number roads connected to the grid cell, (g) the number of cell towers, (h) social connectedness, (I) the area of greenspace, (j) road length, (k) internet speed, and (l) education level. The other three MSAs are shown in Supplementary Figure 6 to 8. The "reciprocal" refers to the inverse of the spatial or infrastructural distance between two grid cells (i.e., 1 / distance), used as a proxy for connection strength in our modeling framework.

## Changes in resilience level under urban development scenario

Urban development and growth patterns reshape the spatial profile of community resilience. *Resili-Net* could be implemented to generate a community's resilience level for different urban development scenarios by changing the input value of community socio-technical features. The following urban development scenario provides a concrete illustration of how *Resili-Net* handles interactions among indicators by altering a subset of features and tracing their combined effect on resilience levels. In the illustrative scenario, every grid cell labelled as low resilience receives a 20 % boost in (i) the number of hospitals reachable within a 30-minute drive and (ii) total road length, representing targeted infrastructure upgrades typical of urban development. All other indicators remain unchanged, and *Resili-Net* is retrained with the same hyper-parameters to produce the updated resilience map (Fig. 4).

The four MSAs exhibit heterogeneous patterns in how community resilience shifts under the urban development scenario. In Greater Los Angeles, around 33 % of grid cells rise at least one resilience tier (about 12 % climb two tiers or more) and fewer than 5 % decline. This indicates that strategies focusing on healthcare accessibility and transportation infrastructure may be particularly effective in these areas. In contrast, the Greater Chicago area presents a more complex picture. While some communities (18 % of cells) would see an improvement in resilience, others would experience a decline (16 %). This suggests that a one-size-fits-all approach would not be applicable, necessitating more nuanced, localized strategies. Moreover, it implies that factors other than hospital accessibility and road length, such as social cohesion, could be essential and require separate interventions. Within the MSAs for Greater Dallas and Houston, most communities' resilience levels remain unchanged under further urban development, but communities with improved or deteriorated resilience are dispersed across a city (10%). This result suggests that mere infrastructure development, such as hospitals and roads, might not be sufficient to improve resilience universally in all areas of a city. Rather, attention should be given to enhancing social infrastructure, such as green spaces, especially in areas already abundant in physical infrastructure.

This heterogeneous community response is due to the intricate interactions among components of socio-technical systems within urban environments, such as social, facility, and infrastructural systems. This complexity underscores the necessity of capturing these interactions to develop targeted resilience enhancement strategies. For instance, communities with already robust and redundant road networks and hospital accessibility may not benefit substantially from further improvements in these domains. Shifting focus towards enhancing social infrastructure, (e.g., green spaces) can simultaneously strengthen well-being and social cohesion, thereby advancing community resilience in multifaceted ways. These strategies could be informed by a deep understanding of the multifaceted interactions among community socio-technical systems which has already been captured by the *Resili-Net* model. Although we present one combined scenario here, *Resili-Net* can readily be rerun with alternative single-indicator or social-infrastructure scenarios in subsequent studies.

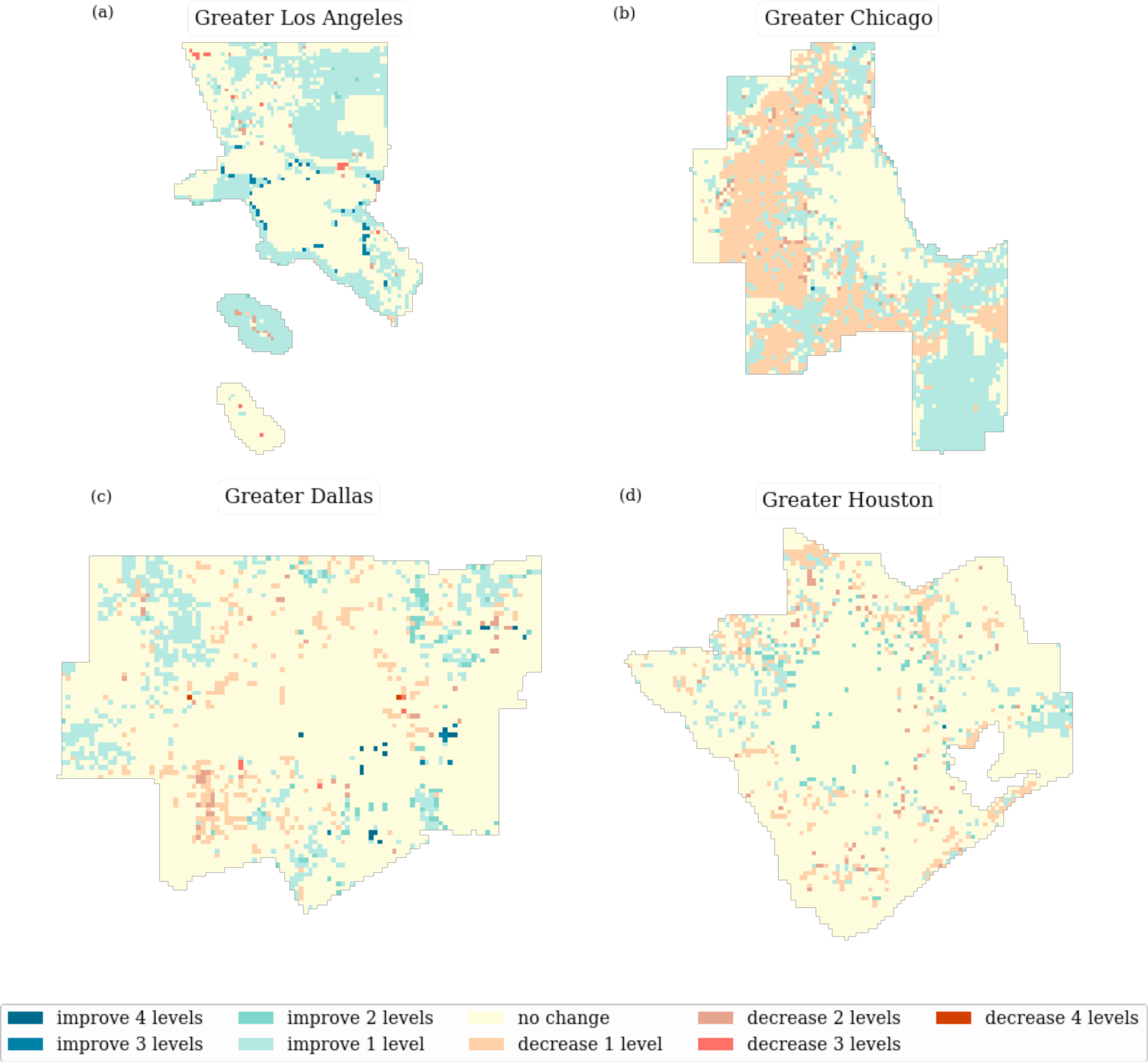

**Fig.4 | Spatial distribution of community resilience level changes under urban development scenario.** Each community's resilience level after urban development is compared with the previous resilience level. Urban development is modeled by the changes in the input feature value. The spatial distribution of community resilience level after urban development is shown in Supplementary Figure 9. The urban development scenario refers to a targeted infrastructure-upgrade scenario applied to grid cells in the lowest resilience tier: each of those cells receives a 20 % increase in (i) the number of hospitals reachable within a 30-minute drive and (ii) total road length, while all other indicators remain unchanged.

### Combined resilience-risk levels

Community resilience levels capture the ability of a community to prepare for, absorb, recover from, and adapt to disturbances[2, 23], while the risk is generally defined as an aggregate of hazard, exposure and vulnerability[24, 25]. Hence, a combined resilience-risk assessment would provide a more comprehensive understanding of the extent of risks in a community and the ability of the community to cope with the risks. In this step, an analysis of the combined resilience and risk levels across four MSAs informs risk mitigation and resilience enhancement plans and actions. The flood risk level of areas is computed using the *FloodRisk-Net* model developed by Yin and Mostafavi [19]. *FloodRisk-Net* captures features of hazards, exposure, and social and physical vulnerability in specifying flood risk levels of spatial areas. Although there is some overlap in input features (building age and poverty rate) in both *Resili-Net* and *FloodRisk-Net*, their functional interpretations differ. In *FloodRisk-Net*, these variables contribute to vulnerability and thus amplify flood risk. In *Resili-Net*, they reflect structural or social capacities that shape resilience (More details on *FloodRisk-Net* is provided in Supplementary Note 1). We use flood risk as an example to perform the combined resilience-risk assessment, while the framework is flexible and can be adapted to incorporate risk levels from other hazards, enabling similar analyses across different threat contexts.

The analysis reveals areas suffering from high flood risk and low resilience levels (Fig. 5). These areas are characterized as lack of infrastructure robustness and redundancy, as well as greater social vulnerability and low resourcefulness of the populations. The high-risk and low-resilience areas are where socially vulnerable populations reside with greater rates of unemployment, disability, elderly, and children. All of these features make these communities susceptible to disasters with limited ability to prepare for, absorb, recover from, and adapt to disturbances. These are areas where infrastructure investments can be optimized to shift resources to other high-risk, low-resilience areas. This requires a nuanced approach to ensure that reducing the allocation of resources does not unintentionally decrease the resilience level.

Analyses of characteristics of areas in high-high and low-low categories (Fig. 6) show that the effects of facilities and infrastructures on community resilience and flood risk are twofold. The large available number of infrastructures and facilities increase the resourcefulness and redundancy of these communities, which is beneficial to the communities' resilience. These extensive physical infrastructure and facility developments would also increase physical vulnerability which increases the flood risk level of the community. Another interesting observation is that in the low-low areas, the low level of resilience may not be a pressing concern due to the relatively low flood-hazard exposure. Nevertheless, while the focus on these areas may not be urgent, it should not be entirely ignored. Infrastructure in these areas should be gradually improved to ensure basic robustness and redundancy, preparing for unforeseen flood risks that may manifest due to climate change or other variables. For high–high areas, where communities already exhibit infrastructure, redundant systems and high resourcefulness, further expansion of traditional infrastructure may not be the most effective strategy. Intensifying physical development in such zones can inadvertently increase exposure to hazards without significantly improving resilience. Instead, these areas would benefit more from risk mitigation strategies, such as low-impact development (e.g., green infrastructure, permeable surfaces, and stormwater management), which reduce hazard risk while reinforcing existing resilience capacities. In parallel, targeted efforts to reduce social vulnerability, such as improving access to healthcare, and community support services, can further mitigate the risk of these communities. This integrated approach promotes sustainable development tailored to the unique needs and risks of these areas.

To further enhance resilience, integrated urban design strategies should be specifically designed for the different types of areas. In high-low areas, retrofitting existing infrastructure to maintain robustness could be a viable strategy. High-low areas could serve as places for establishment of community centers (e.g., shelters) to enhance the overall community resilience. The high social vulnerability in low-high areas with a large percentage of poverty contributes to low robustness and resourcefulness in these areas. To enhance their robustness, infrastructure development strategies should focus on risk reduction and resilience improvement simultaneously. By focusing on these aspects, communities can develop targeted and effective resilience enhancement and risk reduction strategies that account for co-benefits and trade-offs among socio-technical features shaping the varying degrees of flood risk and resilience in different areas. This comprehensive approach ensures that efforts to enhance community resilience are equitable and effective in the long term.

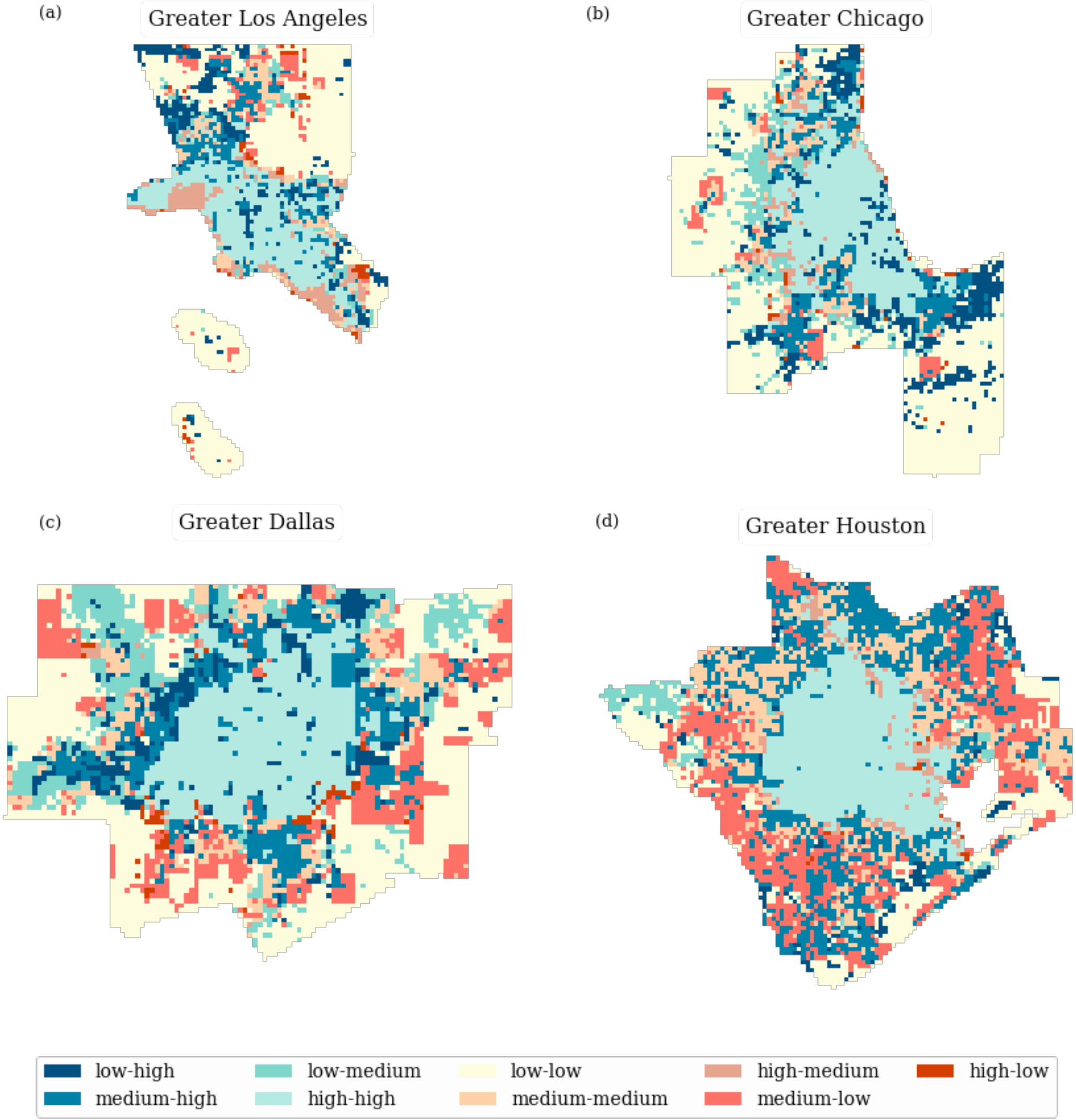

**Fig. 5 | Spatial distribution of the overlap of community resilience and risk level.** Both community resilience level and flood risk level are divided into three types: low, medium, and high. For community resilience: levels 1 and 2 are considered low; level 3, medium; levels 4 and 5, high. For flood risk: levels 1 and 2 are considered low; levels 3 and 4, are medium; and levels 5 and 6, are high. Low-high represents low flood risk level and high community resilience level, the other eight annotations follow the same pattern; that is, the first item represents flood risk level, and the second one is the community resilience level.

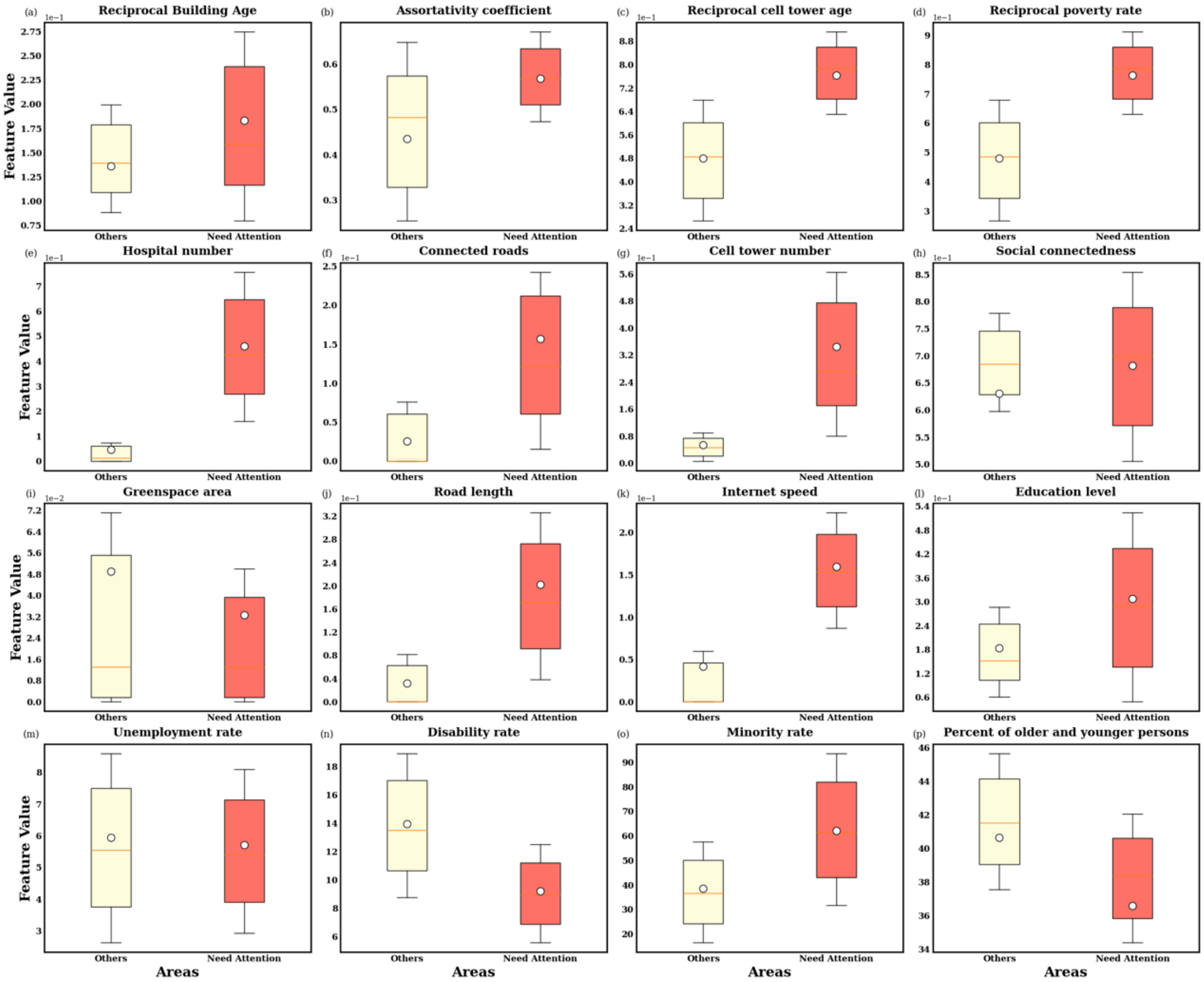

**Fig.6 | Comparisons between characteristics of clusters in areas that need special attention and other areas in Greater Houston**. Areas needing special attention are characterized as high-low, high-medium, and medium-low in Fig. 6 according to their risk and resilience level. The last four of these sixteen features are selected to further evaluate the social vulnerability of communities following Yin and Mostafavi's work [19]. These four features are calculated using 2020 Centers for Disease Control/Agency for Toxic Substances and Disease Registry Social Vulnerability Index (SVI) dataset at the Census Tract level [26]. Unemployment rate represents percentage of the civilian (non-institutionalized) population aged 16 years and older who are unemployed and actively seeking employment. Disability rate represents percentage of the civilian (non-institutionalized) population reporting any hearing, vision, cognitive, ambulatory, self-care, or independent-living difficulty. Minority rate represents percentage of persons not classified as "White, non-Hispanic" (i.e., 100 % minus percent non-Hispanic White). Percentage of older and younger persons is the rate of persons aged 17 and younger and persons aged 65 and older. (The comparisons for the other three MSAs are shown in Supplementary Figure 10 to 12).

## Discussion

This study develops a novel deep learning model, *Resili-Net*, to examine the community resilience profile of cities based on their intertwined socio-technical systems features. *Resili-Net* contains three modules. The first module encodes interactions among twelve resilience-related features spanning three key urban sub-systems: facility, infrastructure, and society. This module employs an encoder-decoder neural architecture to generate low-dimensional latent representations that reflect the embedded, nonlinear structure of resilience dynamics across grid cells. The second module leverages deep embedded clustering to group grid cells into discrete resilience categories based on their learned latent features. Each cluster corresponds to a distinct resilience group, capturing shared structural and social characteristics related to robustness, redundancy, and resourcefulness. The third module then interprets the resilience grouping through an explainable machine learning model, showing the relative importance of each feature in shaping community resilience. These feature importance scores are then used as weights to determine the community resilience level for each group. *Resili-Net* model is applied across four MSAs in the U.S., demonstrating its ability to generalize across diverse urban context. By juxtaposing hazard risk levels and community resilience ratings, we identified critical zones that face both high flood risk and low resilience, enabling priority targets for risk reduction efforts and resilience enhancement strategies.

The findings provide multiple important contributions to the community resilience assessments. First, this study introduces a novel, data-driven approach to assessing community resilience through an integrated unsupervised deep learning architecture. This approach captures the nonlinear and complex interdependencies among multiple features derived from urban socio-technical systems. Through node representation learning, the model encodes latent structures reflecting how these diverse features interact to shape resilience. Deep embedded clustering then groups spatial units into distinct resilience categories based on these learned patterns, revealing emergent community resilience profiles that are difficult to capture using conventional methods. This approach models community resilience as an emergent property of interconnected urban systems, enabling a more nuanced understanding of robustness, redundancy, and resourcefulness at the community scale. While traditional indicator-based methods remain widely used due to their transparency and ease of implementation, they may face limitations when it comes to capturing the full complexity of interacting features. For example, Li and Mostafavi [27] found that using indicator-based methods, which apply equal weights and ignore interactions between urban features can lead to unintended mischaracterizations in infrastructure provision assessment. In contrast, *Resili-Net* provides a complementary, data-driven alternative that automatically learns these relationships from the data, enabling a more context-sensitive and scalable resilience evaluation. This framework is applicable across diverse urban settings using publicly available data, offering researchers and practitioners a powerful tool to better understand, monitor, and compare resilience across communities.

Second, this study bridges the gap between complex deep learning outputs and practical decision-making by providing interpretable resilience level ratings for communities. While deep learning-based clustering is powerful in capturing nonlinear and high-dimensional patterns, it often lacks transparency, making it difficult for practitioners to understand why certain areas are grouped together or how resilience levels are determined. To address this, we implement a post hoc interpretability approach using a random forest classifier trained on the original features and clustering labels. This allows us to derive feature importance scores, which are then used to calculate weighted resilience scores for each cluster. The result is a transparent, data-informed rating of community resilience levels that supports more targeted and actionable planning decisions. Beyond feature importance, this study adds another dimension of interpretability through scenario-based analysis, which enhances understanding of how resilience levels respond to changes in urban development. By simulating changes in key socio-technical features, such as hospital accessibility or road infrastructure, *Resili-Net* can be used to evaluate how resilience levels shift under different urban development scenarios. This enables stakeholders to explore the potential impacts of policy or investment decisions and better understand how specific interventions may strengthen or weaken community resilience across different urban contexts. By revealing how communities respond differently to the same interventions, scenario analysis offers a deeper understanding of the interactions among urban systems and supports the development of targeted, context-sensitive resilience strategies. The interpretability of *Resili-Net* turns resilience modeling into a tool not only for assessment, but also for forward-looking, action-oriented planning.

Third, this study contributes a combined resilience–risk analysis by integrating community resilience profiling with flood risk mapping. This integrated approach enables the identification of compound vulnerability zones, namely the areas that are both highly exposed to flood risk and structurally low in resilience. By mapping these resilience–risk profiles across metropolitan areas, *Resili-Net* supports spatial prioritization of interventions, allowing planners and decision-makers to target areas that face the most severe challenges and have the greatest need for resilience enhancement. Moreover, the results highlight critical insights for urban development strategy. In high-resilience areas, continued infrastructure expansion may not substantially improve resilience, and can, in some cases, increase hazard exposure, suggesting the need for optimal development thresholds. In contrast, low-resilience, low-risk areas may benefit from infrastructure investment to promote long-term resilience and economic opportunity. Meanwhile, high-resilience, low-risk zones can serve as support hubs in times of disaster, helping nearby vulnerable areas through resource diffusion. This spatial differentiation emphasizes the importance of aligning resilience-building efforts with hazard risk contexts, rather than relying on uniform strategies across all areas. *Resili-Net* complements existing methods by offering a dual perspective on hazard and resilience. It equips public officials, planners, and disaster researchers with a more comprehensive understanding of where risk reduction and resilience building efforts should be focused, supporting more strategic, context-sensitive planning and investment.

While this study offers insights into community resilience assessment, there are areas where future research could extend and refine the *Resili-Net* framework to further enhance its scope and applicability. First, while this study conceptualizes community resilience through the lens of socio-technical systems, it does not yet account for the growing recognition of ecological dynamics as a critical component of urban resilience. Emerging research increasingly advocates for a Social-Ecological-Technological Systems (SETS) perspective to capture the co-evolving relationships and feedback loops among natural ecosystems, built infrastructure, and human communities, particularly under climate-induced hazards and compounding stressors [14, 28, 29].Expanding the *Resili-Net* framework to incorporate ecological variables could further enhance its explanatory power and applicability. Second, this study focuses on three widely recognized dimensions of community resilience, robustness, redundancy, and resourcefulness, but does not include rapidity, the fourth dimension in the commonly referenced 4Rs framework. While rapidity is an important characteristic reflecting the speed at which a system can recover from disruption, we were unable to incorporate this dimension due to the lack of spatially consistent and reliable data across all four metropolitan areas analyzed. As such, we limited our analysis to the dimensions for which sufficient data were available. As part of our ongoing efforts to expand the framework, future research will aim to include rapidity by identifying or generating suitable data sources that can support a more complete assessment of community resilience across all four dimensions. Third, this study uses Meta's Social Connectedness Index (SCI) to represent societal redundancy, leveraging its fine-grained spatial detail and broad coverage of social ties. While SCI provides a proxy for social connectivity, it is based on Facebook interactions and may not fully capture the diversity of social networks across all demographic groups. We recognize this as a potential limitation and see future opportunities to complement SCI with additional data sources that reflect a wider range of social interactions and community structures.

## Methods

### Selection of community resilience features

The community is the emergent property arising from complex interactions among several urban sub-systems such as facilities, infrastructures, and society (e.g., transportation system, power system, and communication network) [30]. Drawing on prior community-resilience research (particularly flood-focused studies), we construct our feature set around three core dimensions: robustness, redundancy, and resourcefulness, operationalizing each across the study's facility, infrastructure, and social sub-systems [10, 31, 32].

- **Measurement of robustness**

Robustness denotes the level of performance that can be retained under perturbation [33]. Building age measures the robustness of the facility. As summarized by Yin and Mostafavi [19], older buildings constructed are less advanced than the newly built ones considering their structure materials, construction practices, and building codes; thus older buildings tend to have a higher risk of structure damage under disruptive events [34-36]. The National Structure Inventory (NSI) dataset which provides building age information in the census-tract level was adopted [37]. Following Yin and Mostafavi's work [19], the building age of each grid cell is calculated by the mean value of building age across all census tracts intersected with it, as shown in Equation (1).

$$GC_BA_i = \frac{\sum_j CS_BA_j}{|BA_i|} \; for \; j \; in \; BA_i \tag{1}$$

where, $GC_BA_i$ is the building age of grid cell $i$, $CS_BA_j$ is the building age of census tract $j$, $BA_i$ is the set of census tracts intersected with grid cell $i$, and $||$ is the number of elements in the set (the same hereafter).

Two key infrastructures are considered in our community resilience assessment (i.e., transportation system and communication system) given data availability and their importance in urban systems [38-41]. Assortativity coefficient is adopted to evaluate the robustness of the transportation systems. Assortativity coefficient, which measures the tendency for nodes in a network to connect to other nodes that are similar [42, 43], is a particularly adequate robustness indicator used in network science literature [44]. Simulation results of real-world road networks conducted by Dong et al [22] showed that node degree assortativity has a significant impact on road network robustness, and the correlation between node degree assortativity and road network robustness is positive. Dataset from OpenStreetMap is converted into a road graph by treating intersections as nodes and road segments as edges. Following Yin et al.'s work [6], we include ten road classes and then dissolve all nodes whose degree equals 2, because these points lie on uninterrupted stretches of road and do not represent actual junctions. The resulting network balances computational efficiency with fidelity to real-world connectivity. This road network will also be used to calculate the redundancy and resourcefulness of the transportation system, and the travel distance between each grid cell, which is needed in calculating the redundancy and the resourcefulness of the facility. Road segments within each grid cell are firstly obtained, and then constructed as a road network. The assortativity of the road network within each grid cell is further calculated using Equation (2) [42, 44].

$$GC_AC_i = \frac{1}{\sigma_q^2} \sum_{jk} jk \left(e_{jk} - q_j q_k\right) \tag{2}$$

where, $GC_AC_i$ is the value of assortativity coefficient of grid cell $i$, $q_k$ is the excess distribution and defined as $q_k = \frac{(k+1)p_{k+1}}{\sum_j j p_j}$, $p_k$ is the degree distribution of the network, $e_{jk}$ is the joint probability distribution of a link having degree $j$ on one end and degree $k$ on the other end, and $\sigma_q^2 = \sum_k k^2 q_k - [\sum_k k q_k]^2$.

Similar to the measurement of building age, the cell tower age is used to measure the robustness of a communication system. Older cell towers are more vulnerable to disruptive events than newly built ones due to its less advanced materials, design, equipped technology, and construction standards [19]. The OpenCellID dataset, which provides the precise location information, service range, and the age of a cell tower, is adopted to calculate the cell tower age of each grid cell. The area served by each cell tower is modeled as a circle with the location of the cell tower as the center of the circle and the service range as the radius of this circle. Grid cells intersected with service circle of the cell tower are then assumed to be served by this cell tower. The cell tower age of each grid cell is calculated by taking the average value of ages of all cell towers served by this grid cell, as shown in Equation (3).

$$GC_CAG_i = \frac{\sum_j C_CA_j}{|CA_i|} \ for\ j\ in\ CA_i \tag{3}$$

where, $GC_CA_i$ is the cell tower age of grid cell $i$, $C_CA_j$ is the cell tower age of cell tower $j$, $CA_i$ is the set of cell towers that serve grid cell $i$.

Following Yin and Mostafavi's work [19], the poverty rate in each grid cell is used to measure the robustness in the social system. The 2020 poverty rate dataset in block group level provided by the US Census Bureau is adopted [26]. Mean value of poverty rates in block groups intersected with the grid cell is then calculated as shown in Equation (4).

$$GC_PV_i = \frac{\sum_j BG_PV_j}{|PV_i|} \ for\ j\ in\ PV_i \tag{4}$$

where, $GC_PV_i$ is the poverty rate of grid cell $i$, $BG_BR_j$ is the poverty rate of block group $j$ intersected with grid cell $i$, and $PV_i$ is the set of block groups which are intersected with grid cell $i$.

- **Measurement of redundancy**

Redundancy is defined as the availability of substitutable components with similar functions in the urban system to enhance its adaptive capacity and ability to absorb shocks, give it reserve capacity for problem solving [31, 45, 46]. Numbers of healthcare facilities within an accessible distance are adopted to measure the redundancy of facilities. Healthcare facilities provide the most fundamental functions of this society to secure the basic life need of residents. The high-resolution geographic location information of healthcare centers was obtained from POI data starting with NAICS code of 621 (ambulatory health care services), 622 (hospitals), and 623 (nursing and residential care facilities). The serviced grid cells of each hospital are determined as grid cells which are within 30 minutes driving time to the hospital. The threshold of 30-minute driving time is selected because it is a frequently adopted threshold to determine the serviced patients of the hospital [47-50]. The estimated shortest distance between grid cells is calculated following the work of Fan et al [51]. The travel time between grid cells is estimated assuming the travel speed of each road segment is its free-flow speed. When one key facility is disrupted within the accessible distance, residents could turn to others, ensuring their basic life needs. The abundance of healthcare facilities could ensure the availability of substitutable components which crucial for preserving life. With travel time information, for each grid cell, the number of available healthcare centers within 30-minutes driving time could be calculated as shown in Equation (5).

$$HG_i = |\{C_j | d(G_i, C_j) \leq 30\}| \tag{5}$$

where, $HG_i$ is the accessible number of healthcare centers of grid cell $i$, $C_j$ is the healthcare center $j$, and $d(G_i, C_j)$ is the driving time from grid cell $i$ to healthcare center $j$.

The number of roads connected with the grid cell is used to measure the redundancy of transportation. The number of roads connected with the grid cell could ensure the travel outside the grid cell to others. When some roads connected with the grid cell are disrupted, alternative roads could be used to ensure the connection between other grid cells. The larger this index is, the larger the redundancy of the transportation network is to ensure basic mobility (e.g., the movement of rescue resources) among different areas. Links and nodes within each grid cell are obtained and the feature is calculated using Equation (6).

$$RI_k = \sum_i RS_i \tag{6}$$

where, $RI_k$ is the number of road segments intersected with the boundary of grid cell $k$, and $RS_i$ is a binary variable which equals 1 if it is intersected with the boundary of $k$, otherwise, 0.

The number of available cell towers is treated to measure the redundancy of the communication system. If more than one cell tower is available to the grid cell, when one of the cell towers is disrupted, others could provide alternatives for residences in this grid cell to maintain communication with the outside world. This feature is important for coordinating rescue and relief operations after a disaster, ensuring that accurate information can be sent and received by affected community despite the challenging circumstances. It will also serve to reassure people affected by the disaster that help is on the way and their situation is known to the outside world. Similar to the calculation of cell tower age, the OpenCellID dataset is also used. The available number of cell towers of each grid cell is calculated as shown in Equation (7).

$$GC_CT_i = |\{C_CT_j | C_CT_j \cap GC_i\}| \quad (7)$$

where, $GC_CT_i$ is the available number of cell tower in grid cell $i$, $C_CT_j$ is the service circle of the cell tower $j$, and $GC_i$ is grid cell $i$.

The Social Connectedness Index (SCI), which measures the extent to which people with low and high socioeconomic status are friends with each other, is adopted to measure the redundancy of the society [52]. Residents with dissimilar hazard exposure would help each other through social ties. Those with wider social connections could have more choices to seek assistance to lessen impacts of disruption. That is, when some of their social connections are unable to help them (may be caused by their unwillingness, their inability, etc.), they could seek help from other social connections. This redundancy in social ties could help residents better cope with disruptions and recover more quickly from disruptions. Connectedness statistics at the Zip code level publicly available from Social Capital Atlas are adopted to calculate the social connectedness index of each grid cell [52]. Each grid cell's SCI value is calculated as the mean value of intersected Zip codes' SCI as shown in Equation (8):

$$GC_SCI_i = \frac{\sum_j ZC_SCI_j}{|Z_i|} \; for \; j \; in \; Z_i \quad (8)$$

where, $GC_SCI_i$ is the SCI value of grid cell $i$, $Z_i$ is a set of Zip code intersected with grid cell, $|Z_i|$ is the number of elements in set $Z_i$, and $ZC_SCI_j$ is the SCI value of the $j^{tZh}$ Zip code intersected with grid cell $i$.

- **Measurement of resourcefulness**

Resourcefulness is defined as the ability to effectively use available resources, both internal and external, to adapt, overcome, or mitigate the impacts of disturbances [30]. After disturbances, green space could provide the spare capacity for temporary shelters to accommodate large numbers of affected residents, and space for building reconstruction [53], which could help the community cope with, recover from disturbances better. Land cover data is frequently used to measure the physical and biological cover over the surface of the earth [54] and National Land Cover Database (NLCD) is adopted to calculate the area of green space of each grid cell. The area of land use type 71 (grassland/herbaceous) in each grid cell is calculated as its green space area.

$$GS_BC_i = \sum\nolimits_j G_BC_j \; for \; j \; in \; GS_j \quad (9)$$

where, $GS_BC_i$ is the green space area of grid cell $i$, $G_BC_j$ is the intersected area of NLCD pixel in type 71 cell $j$ with grid cell $i$, and $GS_j$ is the set of NLCD pixels intersected with grid cell $i$.

Road density is used to measure the resourcefulness of the transportation system. A higher road density can improve access to important locations, such as evacuation facilities and hospitals [55]. This is particularly vital during emergencies when rapid response and evacuation are necessary. Thus, in the face of disturbances, the denser road network would facilitate evacuation of residents and would ensure rescue resources reach affected communities quickly [56, 57]. The total length of road segments inside each grid cell is calculated, and it is further divided by the area of grid cell to obtain the value of road density.

Internet speed measures the resourcefulness of a communication system. When facing disruptive events, faster internet speed ensures that communities have access to real-time, reliable information which are crucial in helping them respond to and recover from disruptive events, as well as ensure effective coordination among emergency services. The internet speed dataset is from Ookla, which is publicly accessible in the resolution of zoom level 16 Web Mercator Tiles (in the shape of pixel). For each grid cell, its internet speed is calculated as the average internet speed of all pixels that intersect with it, as shown in Equation (10).

$$GC_IS_i = \frac{\sum_j P_IS_j}{|IS_i|} \; for \; j \; in \; IS_i \tag{10}$$

where, $GC_IS_i$ is the internet speed of grid cell $i$, $P_IS_j$ is the internet speed of pixel $j$, and $IS_i$ is the set of pixels intersected with grid cell $i$.

Education level is adopted to measure the resourcefulness of society from the resident's ability to apply available resources. Higher education level usually results in better comprehension of data and information related to potential hazards and more effective emergency response. Education could also increase a person's ability to access, understand, and utilize information, and improve critical thinking and problem-solving abilities, which are all critical in making effective strategies in the face of disturbances. The 2020 bachelor's degree rate dataset at the census block level provided by US Census Bureau is adopted [26]. Mean value of bachelor rates in block groups intersected with the grid cell is calculated as the education level of the corresponding grid cell, as shown in Equation (11).

$$GC_EL_i = \frac{\sum_j BG_BR_j}{|BR_i|} \; for \; j \; in \; BR_i \tag{11}$$

where, $GC_EL_i$ is the education level of grid cell $i$, $BG_BR_j$ is the bachelor rate of block group $j$ intersected with grid cell $i$, and $BR_i$ is the set of block groups which are intersected with grid cell $i$.

**Community resilience feature matrix construction**

Twelve resilience-related features pertinent to the three resilience components (robustness, redundancy, and resourcefulness) in urban sub-systems of facilities, infrastructure, and societies (shown in Table 2) are computed to construct the resilience feature matrix ($RF \in R^{m*16}$ being the number of grid cells in each MSA). These resilience-related features are at scales ranging from census-tract level to highly-detailed geographic coordinates and min–max normalized to the 0–1 range. The study area is divided into 2 km ×2 km grid cells respecting both the computation cost and the need for proper scale of analysis related to community resilience profiles. Community resilience-related features are aggregated at the grid-cell level. Each column in $RF$ represents a resilience-related feature, and every row records resilience features for a grid cell. This means we calculate the value of each resilience-related feature at grid cell level. To address potential data quality issues such as missing data, spatial resolution mismatches, or biases, all features are first aggregated to a uniform 2km × 2km grid, and sparse or incomplete grid cells are excluded from training to ensure feature consistency and reliability. Grid cells lacking minimum data density (e.g., no cell-tower records) are excluded from training, ensuring that only features with reliable support enter the model. The data sources of our community resilience assessment features are summarized in Supplementary Tabel 1.

**Table 2 Features used to assess community resilience**

| Urban Sub-systems | Robustness | Redundancy | Resourcefulness |
|---|---|---|---|
| Facility | Building age | Number of hospitals within 30 minutes driving distance | Area of green space |
| Infrastructure-Transportation | Degree assortativity coefficient | Number of roads connected with grid cell | Road length |
| Infrastructure-Communication | Cell tower age | Number of cell towers | Internet speed |
| Society | Poverty rate | Social connectedness | Education level |

**Node representation learning**

The stacked denoising autoencoders (SDAE), an autoencoder model developed by Vincent et al. [58], are deployed to capture complex and nonlinear interactions among individual community resilience-related features by encoding these features into latent representations. In SDAE, the autoencoder is first pre-trained in the sub-autoencoder level. Each sub-autoencoder is a denoising autoencoder which is a neural network that receives a corrupted data as input and is trained to reconstruct the original uncorrupted data as output, which is defined as:

$$\begin{aligned} \tilde{x} &= Dropout(x) \\ h &= f_1(W_1\tilde{x} + b_1) \\ \tilde{h} &= Dropout(h) \\ y &= f_2(W_2\tilde{h} + b_2) \end{aligned} \tag{12}$$

where $x$ is the original uncorrupted data, $\tilde{x}$ is the actual input data of encoder layer obtained after dropout, $f_1$ and $f_2$ are activation functions for the encoding and decoding layer, respectively, and $\{W_1, b_1, W_2, b_2\}$ are trainable model parameters. As we aim to reconstruct the uncorrupted data, the loss function is designed as $||x - y||_2^2$. The rectified linear unit (ReLU) is adopted as the activation function for each sub-autoencoder, except for $f_1$ of the last sub-autoencoder to make the final node embeddings preserve full information [58]. The embedding obtained through the encoder layer is used as the input of next sub-autoencoder to train it.

After greedy layer-wise pre-train, the encoder layer in each sub-autoencoder is concatenated followed by all decoder layers in reverse lay-wise training order [58]. A deep autoencoder is thus developed and then fine-tuned to minimize reconstruction loss [58]. The resilience-feature matrix is passed through the encoder layers of this fine-tuned deep autoencoder ($f_e$) to obtain the encoded representations ($E$).

**Deep embedded clustering**

Inspired from Xie et al.'s work [59], a self-training strategy is employed to jointly optimize the assignment of cluster labels and learn features conductive for clustering. Xie et al. [59] proposed to obtain the final clustering result through alternating between two steps until a convergence criterion is reached. First, a soft assignment between $E$ and the cluster centroids are calculated. Second, a high confidence assignment is learned using an auxiliary target distribution. It is further used to update $f_e$ for more cluster-friendly embeddings and refine cluster centroids.

The standard *k*-means clustering is used once on $E$ to obtain initial cluster centroids $u_j$. Following the work of Maaten and Hinton [60], the S *t*-distribution is used as a kernel to measure the similarity between embedding $z_i$ and $u_j$:

$$q_{ij} = \frac{(1 + ||z_i - u_j||^2/\alpha)^{-\frac{\alpha+1}{2}}}{\sum_{j'}(1 + ||z_i - u_{j'}||^2/\alpha)^{-\frac{\alpha+1}{2}}} \tag{13}$$

where $z_i$ is the $i - th$ sample of $E$, $q_{ij}$ is the probability of assigning sample $i$ to cluster center $j$, and $\alpha$ is the degree of freedom of the *t-distribution* and it is set to be 1 following the work of Xie et al [59].

Following Xie et al. [59], the auxiliary target distribution ($P$) is defined by first raising $q_i$ to the second power and then normalizing by frequency per cluster:

$$p_{ij} = \frac{{q_{ij}}^2/f_j}{\sum_{j'} {q_{ij'}}^2/f_{j'}} \tag{14}$$

where $f_j = \sum_i q_{ij}$ are soft cluster frequencies. The soft assignment ($Q$) is trained to match the auxiliary target distribution ($P$) by minimizing the KL divergence loss between the soft assignment ($q_i$) and the auxiliary target distribution ($p_i$) as follows [59]:

$$L = KL(P||Q) = \sum_i \sum_j p_{ij} \frac{p_{ij}}{q_{ij}} \tag{15}$$

The most likely cluster $k$ that instance $i$ belongs ($s_i$) to is.

$$s_i = \underset{k}{\operatorname{argmax}}(q_{ik}) \tag{16}$$

To test the robustness of our proposed method, we reran the clustering step for the Houston MSA using five different random *k*-means initializations and two alternative values for the *t*- distribution parameter ($\alpha$ = 0.5 and $\alpha$ = 2). Across these runs, fewer than 5 % of grid cells changed their final cluster label, and city-level summary statistics were unaffected, indicating that the model is insensitive to the specific centroid seed or $\alpha$ value.

**Community resilience level determination**

Grid cells in the same group are assumed to have the same community-resilience level due to the clustering module having clustered grid cells with similar community resilience into the same group. The clustering module also ensures that grid cells in different clusters are different from each other as much as possible.

Inspired from the work of Morichetta et al. [61], work, an explainable classification model is implemented to interpret the clustering results. We do this by training an explainable classification model on the clustering results obtained in the second layer of the model, i.e., using the clustering results as the labels of the classification model then interpreting the cluster assignment results modeled by the explainable classifier.

The random forests classifier was selected as the explainable classification model considering its interpretability and performance [21]. Random forests classifier is implemented using Scikit-learn [20] by taking resilience-related feature matrix as input and setting clustering results as labels, and details of random forests classifier is described in the work of Breiman [21].

After obtaining feature importance of individual resilience-related feature to the final clustering results, the community-resilience level is determined as follows:

1. Following Siam et al. [20], Xu et al. [62], and Yin and Mostafavi's [19] work community resilience-related features are normalized to reduce the impact of the difference of unit using Min-Max scaler:

$$x' = \frac{x - x_{min}}{x_{max} - x_{min}} \tag{17}$$

where $x'$ is the scaled value which is in the range of $[0,1]$.

2. Following the work of Yin and Mostafavi work [19], calculate each community resilience-related feature's mean values for each cluster:

$$\overline{R_{iku}} = \frac{\sum_{l=1}^{n} R_{iku_l}}{n} \tag{18}$$

where $\overline{R_{iku}}$ is the average value for feature $i$ in cluster $k$ in MSA $u$, and $R_{iku_l}$ is the $l$-th value of feature $i$ in cluster $k$ in MSA $u$.

3. The aggregated community resilience value of each cluster is calculated by weighted sum of each resilience-related feature by taking feature importance as the weight:

$$AR_{ku} = \sum_{i} IM_i * \overline{R_{iku}} \tag{19}$$

where $IM_i$ is the importance of feature $i$ to the clustering result, and $AR_{ku}$ is the aggregated resilience value of cluster $k$ in MSA $u$.

4. Finally, calculate the flood risk level value $RL_{ku}$ by ranking the value of $AR_{ku}$ in an ascending way:

$$RL_{ku} = Sorted\,(AR_{ku}) \tag{20}$$

**Parameters setting**

We set the searching space for each hyperparameter in *Resili-Net* and use grid search to select the optimal set of hyperparameters. The network dimension of $f_e$ is set as $d_{in} - 500 - 500 - 2000 - d_e$, where $d$ is the input dimension, and $d_e$ is the embedding dimension. The two 500-unit hidden layers and 2000-unit decoder mirror the configuration adopted in *FloodRisk-Net* and, in a pilot run on the Houston MSA, kept reconstruction error below 1 % while fitting comfortably in GPU memory. The embedding dimension $d_e$ is grid-searched over [10, 12, 24, 36]. The Silhouette scores levelled off at 24, so that value is used in the final model. The pre-train and fine-tune epochs are set at 200. The dropout rate is fixed at 0.2, which is widely used in comparable deep embedded clustering implementations and provided stable convergence here. The batch size is set at 256 and the learning rate is set at 0.1. The clustering number is grid searched from [4, 5, 6, 7], with $k = 5$ yielding the best average silhouette score.

The performance of the classifier is evaluated by precision, recall, and F1-score in macro-, micro, and the value of AUC. Definitions of these evaluations can be referenced in the work of Ho et al. [63]. The random forest classifier is capable of modeling, almost entirely, the clustering results, with AUC. F1-score, recall, precision values all close to 1 for all four clustering (Fig. 7). This demonstrates that the classifier is able to represent the obtained cluster assignments as classification classes, and can be automatically used to interpret the allocation of instances to each cluster.

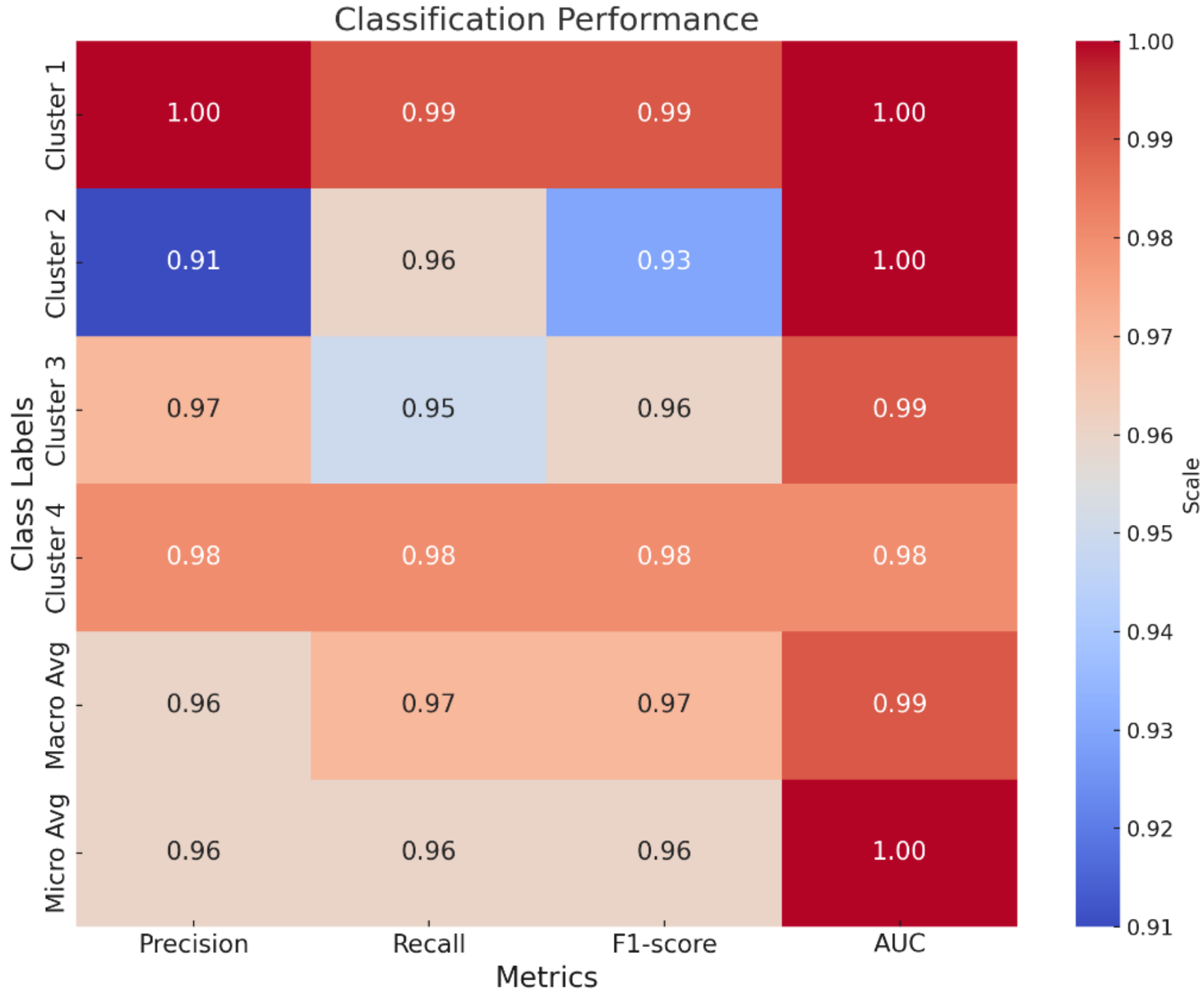

**Fig. 7 | Performances of random forest classifier results with cluster results as labels for Greater Houston.** Other three MSAs are in Supplementary Figure 13 to 15.

**Data Availability**

The dataset used in this paper are publicly accessible and cited in this paper. Users could directly download the needed datasets through the website cited.

**Acknowledgements**

This work was supported by National Science Foundation under CRISP 2.0 Type 2 No. 1832662 grant. Any opinions, findings, conclusions, or recommendations expressed in this research are those of the authors and do not necessarily reflect the view of the funding agencies.

**Author contributions**

**Kai Yin:** Conceptualization, Methodology, Software, Formal analysis, Investigation; Writing– original draft, Writing – Review & Editing, Visualization. **Bo Li**: Conceptualization, Investigation, Writing – Original Draft, Writing – Review & Editing, **Ali Mostafavi:** Conceptualization; Methodology; Writing, Reviewing and Editing; Supervision; Funding acquisition. All authors have read and approved the manuscript.

**Competing interests**

The authors declare no competing interests.

**Additional information**

Supplementary material associated with this article can be found in the attached document.

## References

1. Hallegatte, S., et al., *Future flood losses in major coastal cities.* Nature climate change, 2013. **3**(9): p. 802-806.
2. Academies, N., et al., *Disaster resilience: A national imperative.* 2012: National Academies Press.
3. Esmalian, A., et al., *Operationalizing resilience practices in transportation infrastructure planning and project development.* Transportation Research Part D: Transport and Environment, 2022. **104**: p. 103214.
4. Lee, C.-C., et al., *Quantitative measures for integrating resilience into transportation planning practice: Study in Texas.* Transportation Research Part D: Transport and Environment, 2022. **113**: p. 103496.
5. Rajput, A.A. and A. Mostafavi, *Latent sub-structural resilience mechanisms in temporal human mobility networks during urban flooding.* Scientific Reports, 2023. **13**(1): p. 10953.
6. Yin, K., et al., *An integrated resilience assessment model of urban transportation network: A case study of 40 cities in China.* Transportation Research Part A: Policy and Practice, 2023. **173**: p. 103687.
7. Bhusal, N., et al., *Power system resilience: Current practices, challenges, and future directions.* Ieee Access, 2020. **8**: p. 18064-18086.
8. Wang, Y., et al., *Research on resilience of power systems under natural disasters—A review.* IEEE Transactions on power systems, 2015. **31**(2): p. 1604-1613.
9. Ma, J., et al., *Establishing nationwide power system vulnerability index across US counties using interpretable machine learning.* Applied Energy, 2025. **397**: p. 126360.
10. Saja, A.A., et al., *An inclusive and adaptive framework for measuring social resilience to disasters.* International journal of disaster risk reduction, 2018. **28**: p. 862-873.
11. Ma, J. and A. Mostafavi, *Decoding the pulse of community during disasters: Resilience analysis based on fluctuations in latent lifestyle signatures within human visitation networks.* International Journal of Disaster Risk Reduction, 2025. **124**: p. 105552.
12. Cimellaro, G.P., et al., *PEOPLES: a framework for evaluating resilience.* Journal of Structural Engineering, 2016. **142**(10): p. 04016063.
13. Datola, G., *Implementing urban resilience in urban planning: A comprehensive framework for urban resilience evaluation.* Sustainable Cities and Society, 2023. **98**: p. 104821.
14. Helmrich, A., et al., *Interdependence of social-ecological-technological systems in Phoenix, Arizona: Consequences of an extreme precipitation event.* Journal of Infrastructure Preservation and Resilience, 2023. **4**(1): p. 19.
15. Moghadas, M., et al., *A multi-criteria approach for assessing urban flood resilience in Tehran, Iran.* International journal of disaster risk reduction, 2019. **35**: p. 101069.
16. Anelli, D., F. Tajani, and R. Ranieri, *Urban resilience against natural disasters: Mapping the risk with an innovative indicators-based assessment approach.* Journal of Cleaner Production, 2022. **371**: p. 133496.
17. Parizi, S.M., M. Taleai, and A. Sharifi, *A GIS-based multi-criteria analysis framework to evaluate urban physical resilience against earthquakes.* Sustainability, 2022. **14**(9): p. 5034.
18. Parizi, S.M., M. Taleai, and A. Sharifi, *Integrated methods to determine urban physical resilience characteristics and their interactions.* Natural Hazards, 2021. **109**(1): p. 725-754.
19. Yin, K. and A. Mostafavi, *Unsupervised graph deep learning reveals emergent flood risk profile of urban areas.* arXiv preprint arXiv:2309.14610, 2023.
20. Siam, Z.S., et al., *National-scale flood risk assessment using GIS and remote sensing-based hybridized deep neural network and fuzzy analytic hierarchy process models: a case of Bangladesh.* Geocarto International, 2022. **37**(26): p. 12119-12148.
21. Breiman, L., *Random Forests.* Machine Learning, 2001. **45**: p. 5-32.
22. Dong, S., et al., *Measuring the topological robustness of transportation networks to disaster-induced failures: A percolation approach.* Journal of Infrastructure Systems, 2020.

**26**(2): p. 04020009.
23. NIAC, *Crticial Infrastructure Resilience Final Report and Recommendations*. 2009.
24. Herreros-Cantis, P., et al., *Shifting landscapes of coastal flood risk: environmental (in) justice of urban change, sea level rise, and differential vulnerability in New York City.* Urban transformations, 2020. **2**(1): p. 9.
25. Lavell, A., et al., *Managing the risks of extreme events and disasters to advance climate change adaptation.* A special report of working groups I and II of the intergovernmental panel on climate change (IPCC), 2012. **3**: p. 25-64.
26. US Centers for Disease Control and Prevention. *CDC/ATSDR Social Vulnerability Index 2020*. 2020; Available from: https://www.atsdr.cdc.gov/placeandhealth/svi/index.html.
27. Li, B. and A. Mostafavi, *Incorporating environmental considerations into infrastructure inequality evaluation using interpretable machine learning.* Computers, Environment and Urban Systems, 2025. **120**: p. 102301.
28. Chang, H., et al., *Assessment of urban flood vulnerability using the social-ecological-technological systems framework in six US cities.* Sustainable Cities and Society, 2021. **68**: p. 102786.
29. McPhearson, T., et al., *Radical changes are needed for transformations to a good Anthropocene.* Npj urban sustainability, 2021. **1**(1): p. 5.
30. Bruneau, M., et al., *A framework to quantitatively assess and enhance the seismic resilience of communities.* Earthquake spectra, 2003. **19**(4): p. 733-752.
31. Sharifi, A. and Y. Yamagata, *Principles and criteria for assessing urban energy resilience: A literature review.* Renewable and Sustainable Energy Reviews, 2016. **60**: p. 1654-1677.
32. Tong, P., *Characteristics, dimensions and methods of current assessment for urban resilience to climate-related disasters: A systematic review of the literature.* International Journal of Disaster Risk Reduction, 2021. **60**: p. 102276.
33. Gu, Y., et al., *Performance of transportation network under perturbations: Reliability, vulnerability, and resilience.* Transportation Research Part E: Logistics and Transportation Review, 2020. **133**: p. 101809.
34. Fernandez, P., et al., *A new approach for computing a flood vulnerability index using cluster analysis.* Physics and Chemistry of the Earth, Parts a/b/c, 2016. **94**: p. 47-55.
35. Neubert, M., et al., *The geographic information system-based flood damage simulation model HOWAD.* Journal of Flood Risk Management, 2016. **9**(1): p. 36-49.
36. Hebb, A. and L. Mortsch, *Floods: Mapping vulnerability in the Upper Thames watershed under a changing climate.* Project Report XI, University of Waterloo, 2007: p. 1-53.
37. US Army Corps of Engineers Hydrologic Engineering Center. *NSI Technical Documentation*. 2022; Available from: https://www.hec.usace.army.mil/confluence/nsi/technicalreferences/latest/technical-documentation
38. Wang, W., et al., *Network approach reveals the spatiotemporal influence of traffic on air pollution under COVID-19.* Chaos: An Interdisciplinary Journal of Nonlinear Science, 2022. **32**(4).
39. Qi, H., et al., *Sustainable development-oriented campus bike-sharing site evaluation model: A case study of Henan Polytechnic University.* arXiv preprint arXiv:2308.04454, 2023.
40. Yin, K., et al. *Examining the reasons for the low market share of road passenger transport based express using structural equation modeling.* in *18th COTA International Conference of Transportation Professionals*. 2018. American Society of Civil Engineers Reston, VA.
41. Rajput, A.A., et al., *Revealing critical characteristics of mobility patterns in New York City during the onset of COVID-19 pandemic.* Frontiers in Built Environment, 2022. **7**: p. 654409.
42. Newman, M.E., *Assortative mixing in networks.* Physical review letters, 2002. **89**(20): p.

208701.
43. Newman, M.E., *Mixing patterns in networks.* Physical review E, 2003. **67**(2): p. 026126.
44. Derrible, S. and C. Kennedy, *The complexity and robustness of metro networks.* Physica A: Statistical Mechanics and its Applications, 2010. **389**(17): p. 3678-3691.
45. Sharifi, A. and Y. Yamagata. *Major principles and criteria for development of an urban resilience assessment index*. in *2014 international conference and utility exhibition on green energy for sustainable development (ICUE)*. 2014. IEEE.
46. Tyler, S. and M. Moench, *A framework for urban climate resilience.* Climate and development, 2012. **4**(4): p. 311-326.
47. Shen, Y.C. and R.Y. Hsia, *Differential benefits of cardiac care regionalization based on driving time to percutaneous coronary intervention.* Academic Emergency Medicine, 2021. **28**(5): p. 519-529.
48. Rayburn, W.F., M.E. Richards, and E.C. Elwell, *Drive times to hospitals with perinatal care in the United States.* Obstetrics & Gynecology, 2012. **119**(3): p. 611-616.
49. Chang, J., et al., *Driving time to the nearest percutaneous coronary intervention-capable hospital and the risk of case fatality in patients with acute myocardial infarction in Beijing.* International journal of environmental research and public health, 2023. **20**(4): p. 3166.
50. Koeze, E., J. Patel, and A. Singhvi, *Where Americans live far from the emergency room.* The New York Times, 2020. **26**.
51. Fan, C., et al., *Equality of access and resilience in urban population-facility networks.* npj Urban Sustainability, 2022. **2**(1): p. 9.
52. Chetty, R., et al., *Social capital I: measurement and associations with economic mobility.* Nature, 2022. **608**(7921): p. 108-121.
53. Shrestha, S.R., R. Sliuzas, and M. Kuffer, *Open spaces and risk perception in post-earthquake Kathmandu city.* Applied geography, 2018. **93**: p. 81-91.
54. Liu, J., et al., *From isolation to linkage: Holistic insights into ecological risk induced by land use change.* Land Use Policy, 2024. **140**: p. 107116.
55. Bozza, A., D. Asprone, and G. Manfredi, *Developing an integrated framework to quantify resilience of urban systems against disasters.* Natural Hazards, 2015. **78**(3): p. 1729-1748.
56. Sakakibara, H., Y. Kajitani, and N. Okada, *Road network robustness for avoiding functional isolation in disasters.* Journal of transportation Engineering, 2004. **130**(5): p. 560-567.
57. Kotzee, I. and B. Reyers, *Piloting a social-ecological index for measuring flood resilience: A composite index approach.* Ecological indicators, 2016. **60**: p. 45-53.
58. Vincent, P., et al., *Stacked denoising autoencoders: Learning useful representations in a deep network with a local denoising criterion.* Journal of machine learning research, 2010. **11**(12).
59. Xie, J., R. Girshick, and A. Farhadi. *Unsupervised deep embedding for clustering analysis*. in *International conference on machine learning*. 2016. PMLR.
60. Maaten, L.v.d. and G. Hinton, *Visualizing data using t-SNE.* Journal of machine learning research, 2008. **9**(Nov): p. 2579-2605.
61. Morichetta, A., P. Casas, and M. Mellia. *EXPLAIN-IT: Towards explainable AI for unsupervised network traffic analysis*. in *Proceedings of the 3rd ACM CoNEXT Workshop on Big DAta, Machine Learning and Artificial Intelligence for Data Communication Networks*. 2019.
62. Xu, H., et al., *Urban flooding risk assessment based on an integrated k-means cluster algorithm and improved entropy weight method in the region of Haikou, China.* Journal of hydrology, 2018. **563**: p. 975-986.
63. Ho, Y.-H., T.-C. Hsiao, and A.Y. Chen, *Emission analysis of electric motorcycles and assessment of emission reduction with fleet electrification.* IEEE Transactions on Intelligent Transportation Systems, 2023. **24**(12): p. 15369-15378.